\documentclass[
aps,%
11pt,%
final,%
notitlepage,%
oneside,%
onecolumn,%
nobibnotes,%
nofootinbib,%
superscriptaddress,%
noshowpacs,%
centertags]%
{revtex4}

\usepackage{multirow}

\begin{document}

\selectlanguage{english}
\keywords{galaxies -- dwarf galaxies: star formation}

\title{$H\alpha$ SURVEY OF LOW-MASS SATELLITES OF THE NEIGHBORING
GALAXIES  M31 AND  M81}

{\scriptsize Astrophysics, Vol. 56, No. 3, September , 2013

0571-7256/13/5603-0305 2013 Springer Science+Business Media New York

Original article submitted May 15, 2013.  Translated from Astrofizika, Vol. 56, No. 3, pp. 331-348 (August
2013).}

\author{\firstname{S.~S.}~\surname{Kaisin}}
\author{\firstname{I.~D.}~\surname{Karachentsev}}
\affiliation{Special  Astrophysical  Observatory,  Russian  Academy  of  Sciences, Russia}

\begin{abstract}
Images have been obtained at the 6-m telescope at the Special Astrophysical Observatory (SAO) of the
Russian Academy of Sciences in the $H\alpha$  line and in the continuum for 20 dwarf spheroidal satellites of
M31: And XI-And XXX, plus the distant Globular cluster Bol 520.  Their star formation rates ($FR$) are
estimated using the  $H\alpha$  flux and the ultraviolet $FUV$ flux measured with the GALEX space telescope.
Most of the dSph satellites of M31 have extremely low star formation rates with a characteristic upper
limit of  $SFR\sim 5\times10^{-7}$.  We have made similar estimates of $SFR$ from the  $H\alpha$  and $FUV$ fluxes for 13
galaxies with low surface brightness recently discovered in the neighborhood of M81.  Eleven of them are
physical satellites of M81 with typical   $SFR< 5\times10^{-5}$.  The median stellar masses of these satellites of
M31 and M81 are 0.9 and 1.9 million solar masses, respectively.  Our $H\alpha$  observations place a 2--3 times
stricter limit on the value of $SFR$ than the data from the GALEX satellite, with a substantially higher (3--5
times) angular resolution.

\end{abstract}
\maketitle

\section{Introduction}

Over the last few years the 6-m telescope at the Special Astrophysical Observatory (SAO) of the Russian
Academy of Sciences has been engaged in a program of systematic surveys in the $H\alpha$  line of nearby galaxies in order
to study the features of star formation in them.  Thus far, $H\alpha$  images of more than 300 galaxies at distances within
10 Mpc have been obtained.  The results have been published in a series of papers (Karachentsev et al. 2005, Kaisin \& Karachentsev 2006, 2008,
Karachentsev \& Kaisin 2007, 2010, Kaisin et al. 2011).  The number of galaxies
in the Local volume $ D  < 10$ Mpc) is constantly rising because of large-scale surveys of the sky in the optical and
HI radio (21-cm) line, and already exceeds 800 galaxies (Kaisina et al. 2012, Karachentsev et al. 2013).
Research on nearby galaxies has the obvious advantage
that it can include objects with very low masses and luminosities that lie below the detection threshold at large
distances.

In an earlier paper (Kaisin et al. 2011) we presented the results of an $H\alpha$  survey of 10 faint satellites of the nearest spiral galaxy
M31.  The most recent searches for new satellites of M31 at the 2.5-m INT (Isaac Newton Telescope) revealed about
20 more low luminosity objects, for which the data are summarized elsewhere (McConnachie 2012, McConnachie \& Irvin 2006).  Similar searches for dwarf
galaxies, resolved to stars, have been undertaken in the neighborhoods of another nearby spiral, M81, using the high
quality optics at the 3.5-m CFHT (Canada France Hawaii Telescope) in Hawaii.  This revealed 16 new satellites of
M81 (Chiboucas et al. 2009); their membership in this group was later confirmed by measuring the distances on the Hubble space
telescope (Chiboucas et al. 2013).  The present paper is devoted to a preliminary study of the features of star formation in the low-mass
satellites of M31 and M81.

It should be noted that most of the recently discovered satellites of M31 and M81 with very low surface
brightnesses are dwarf spheroidal (dSph) systems with an old star population, where the gas is exhausted or has been
swept away, i.e., conditions for the formation of new stars do  not exist.  Nevertheless, faint  star formation spotes and
emission objects (Karachentsev \& Kaisin 2007, 2010) have been found in some nearby spheroidal dwarfs, DD044 and KKR25; this was completely
unexpected.  The discovery of these kinds of emission features offers  the possibility of measuring a radial velosity of
the dSph galaxy, which would be difficult to do if it contained  no emitting gas.  In  addition, cases of structureless dE
and dSph-satellites are known (M32 for M31 and KDG61 for M81), which form convenient screens against which faint
emission knots projected from extended outskirts of giant spirals could be seen Karachentsev et al. (2011).

\section{Observations and data analysis}

We obtained CCD images in the $H\alpha$ line and in the continuum for 33 galaxies with low surface brightnesses
and for a single globular cluster Bol520 in the group  M31.  The observations were made between March 2008 and
December 2012 with an average seeing of 1$\farcs0-2\farcs5$; only for a few of the observed objects the seeing was worse than
2$\farcs5$.

All the observations were made on the 6-m telescope at the Special Astrophysical Observatory (SAO) of the
Russian Academy of Sciences on the SCORPIO detector (Afanasiev et al. 2005) with a 2048$\times$2048 pixel matrix in a 2$\times$2 binning mode
and an image scale of 0.18$^{\prime\prime}$/pixel, which yields a full field of view of 6$\farcm1\times6\farcm1$.
Images in $H\alpha$ +[NII] and in the
continuum were obtained by observing the galaxies through a narrow band $H\alpha$ filter  ($\Delta\lambda =75$\AA) with an effective
wavelength  $\lambda-6555$\AA  \ and SED607 (with  $\Delta\lambda=167$\AA,  $\lambda=6063$\AA) and SED707 (with  $\Delta\lambda=207$\AA,
$\lambda=7036$\AA)
intermediate band filters for the continuum.  Typical exposure times for a majority of the galaxies were $2\times$300 s in
the continuum and $2\times600$ s in $H\alpha$.  Since the range of radial velocities in our sample is small, we used the same $H\alpha$
filter for all the observed objects.

A standard procedure for analysis of direct CCD images was used for processing the data. The bias was initially
subtracted from all the data, and then all the images were divided onto a flat field; after that the cosmic spots were
removed and the sky background was subtracted from the images.  The next operation involved comparing all the
images of a given object, after which all the continuum images were normalized to the $H\alpha$  images using 7--19 field
stars and subtracted.

The fluxes were determined for all the $H\alpha$  images with the continuum subtracted using spectrophotometric
standard stars  (Oke 1990) observed on the same night as the objects.  A study of the measurement errors showed that they
are typically $\sim$10--15\%.

\subsection{Results}

{\em Satellites of Andromeda (M31).}   Figure 1 reproduces the mosaic of images of the 20 satellites of M31 and
a single globular cluster Bol520, which lies  at a projected  distance of 110 kpc from Andromeda that is typical of
spheroidal satellites.  The left image in each pair represents the sum of the exposures in $H\alpha$  and the continuum, and
the right image, in the $H\alpha$  line after deduction of the continuum.  The scale of the images and the directions north
and east are indicated by the line segments and arrows in the figures.  The objects are not aligned along the right
ascension, but in order of their names.  With an image quality of about 1$^{\prime\prime}$, some of the dwarf systems are distinctly
resolved into stars (And XIV, And XV, And XVI, And XVIII, and And XXIX).  The angular diameters of the most
extended objects (And XIX, And XXI, And XXII, And XXIII) are somewhat greater than the image frames.

Almost all of the images in $H\alpha$  with the continuum subtracted contain residual blurs of the stars caused by
differences in the image quality, saturation for the bright stars, or an anomalous color index for some.  For this reason,
estimates of the integrated $H\alpha$  flux for the diffuse dwarfs resolved into stars may have large errors. Nevertheless,
within the optical boundaries of the galaxies And I, And XIV, And XVII, And XXII, And XXIII, and And XXIX, we
suspected the existence of compact emission features which are encircled in the images on the right.  Some of these,
as in the case of KKR25 (Makarov et al. 2012), may turn out to be planetary nebulae. Note that an emission filament, probably
originating in the galaxy, is projected onto the image of And XIX.  Thus, for most of the dwarf dSph satellites of
Andromeda, we can only set an upper limit for the flux in the $H\alpha$  line on the basis of our observations.

{\em Satellites of M81.}  We have given a summary of the $H\alpha$  fluxes for all the known satellites of M81 at the time
in Karachentsev \& Kaisin (2007).  In addition, we have discussed (Roychowdhury 2012) the features of star formation in three blue compact dwarfs (BCD) that
were discovered later, drawing on data on the distribution of neutral hydrogen in them based on high-angular resolution
observations at the Indian GMRT (Giant Metrewave Radio Telescope) radio telescope.  Here we provide some new
data for another 13 fainter dwarf systems in the neighborhood of M81.

Figure 2 shows the mosaic of images in the $H\alpha$  line and the continuum for 11 physical satellites of M81 and
two dwarf galaxies (d1009+70 and d1019+69) which appear to be far background objects.  As for the satellites of M31,
the left images in each pair correspond to the sum of the images in $H\alpha$  and the continuum, and the right images are
in $H\alpha$  with the continuum subtracted. The scale and orientation of the images are indicated as before.

These data show that of the 11 physical satellites of M81, compact HII regions are visible in only one,
d0959+68.  This irregularly structured object has been discovered independently (de Mello et al. 2008) and has been designated Clump
II, as a node in a hydrogen bridge joining M81 with the neighboring galaxy NGC 3077.  No signs of $H\alpha$  emission
were detected in the other 10 dwarf satellites with a low surface brightness.

The dwarf system d1019+69 contains a single HII region at its northern edge. We have obtained a spectrum
of this feature with the 6-m telescope at the SAO of the Russian Academy of Sciences and found its radial velocity
to be  $V_h = (557\pm38)$ km/s, which implies a distance estimate of  $D = 9.6$  Mpc.

\section{Star formation rate in low-mass satellites of nearby spirals}

For each galaxy shown in the mosaics of Figs. 1 and 2 we determined the integrated flux (erg/cm 2s) in the $H\alpha$
line  $F_c(H\alpha)$  or an observed upper limit for it, corrected for optical  extinction in the Galaxy  (Schlegel 1998).  We have neglected
internal extinction within a dwarf galaxy, as well as  the contribution of the neighboring [NII] emission doublet, since
both effects are small for dwarf systems.  In accordance with Kennicutt (1998), we found the integrated rate of star formation
in a galaxy using the linear relation

$$\log[SFR]=\log F_c(H\alpha)+2\log D+8.98, \eqno(1)$$
where  $D$ is the distance to the galaxy in Mpc.

Another, independent way of estimating the star formation rate in a galaxy is based on its flux in the far
ultraviolet, $F(FUV$), at a wavelength of  $\lambda_{eff} = 1539$\AA \ in a band with  FWHM = 269\AA.  For most of the galaxies
considered here, these fluxes were measured by the GALEX space  telescope (Gil de Paz et al. 2003, 2007) and are  given in the NED data
base  NED (http://nedwww.ipav.caltech.edu).   Including the correction for the $FUV$ flux owing to extinction in our
Galaxy, the star formation rate is given by Lee et al. (2011).

$$\log[SFR]=\log F_c(FUV)+2\log D-6.78, \eqno(2)$$
where the  $UV$ flux is measured in mJy and the distance  $D$, in Mpc.

Summaries of data on the satellites of M31 and M81 that we have observed are given in Tables 1 and 2,
respectively.  The columns of both tables list the following: (1) galaxy name; (2) its distance in Mpc; (3) morphological
type according to the scheme described in the UNGC catalog (Karachentsev et al. 2013), where dwarf systems are distinguished according
to their shapes as spheroidal (Sph), irregular (Ir), or intermediate (Tr), and according to their mean surface brightness
class as high (H), normal  (N), low (L), and extremely low (X); (4) absolute  $B$ magnitude of the galaxy; (5) logarithm
of the galaxy stellar mass in  solar units estimated from its  $K$ luminosity; (6) logarithm of its hydrogen mass in solar
masses; (7, 8) logarithms of the integrated rate of star formation determined from the $H\alpha$  flux and the $FUV$ flux,
respectively; (9, 10) logarithm of the specific star formation rate relative to unit stellar mass and normalized to the
cosmological time scale of $T_0= 13.7\times10^9$   years, with

$$P=\log(SFR\cdot T_0/M_*). \eqno(3)$$

\begin{table}

\caption{Integral Parameters of the Observed Satellites of M31}
\begin{tabular}{|l|c|c|c|c|c|r|r|r|r|}
\hline
Satelite    & $D$, Mpc &Type& $M_B$& $\log M_*$& $\log M_{HI}$& $\log SFR_{H\alpha}$& $\log SFR_{FUV}$ & $P(H\alpha)$ &$P(FUV)$\\
\hline
AndXI       & 0.73& Sph-X& --6.2& 5.42& <5.53 &  5.75   &   --5.71   & --1.04   & --1.00\\
AndXII      & 0.83& Sph-X& --6.4& 5.51 &<5.64 &$<-$6.43  &   $<-$6.20  & $<-$1.81 &  $<-$1.58\\
AndXIII     & 0.84& Sph-X& --6.8& 5.66 &<5.65& $<-$6.42   &  $<-$6.30   &$<-$1.95  & $<-$1.83\\
AndXIV      & 0.73& Sph-X& --7.7& 6.03 &<5.53 &  -5.82  &  $<-$6.49  &  --1.71  & $<-$2.38\\
AndXV       & 0.76& Sph-L& --8.7& 6.43 &<5.56 &$<-$6.59   &     --    & $<-$2.89  &    --\\
AndXVI      & 0.52& Sph-L& --8.2& 6.23 &<5.24& $<-$6.76   &   --5.81  & $<-$2.86  &  --1.91\\
AndXVII     & 0.74& Sph-X& --7.0& 5.74 &<5.54&  --5.57   &  --5.71   & --1.18   & --1.32\\
AndXVIII    & 1.36& Sph-L& --9.1& 6.60 &--    & $<-$5.94   & $<-$5.81  & $<-$2.40  & $<-$2.27\\
AndXIX      & 0.93& Sph-X& --8.3& 6.28 &--    & $<-$6.24   & $<-$6.27  & $<-$2.39   &$<-$2.42\\
AndXX       & 0.80& Sph-X& --5.8& 5.26& --     &$<-$6.34   &  -5.96  & $<-$1.47  &  --1.09\\
AndXXI      & 0.86& Sph-X& --9.3& 6.66 &--    & $<-$6.31   & $<-$6.24  & $<-$2.83  & $<-$2.76\\
AndXXII     & 0.79& Sph-X& --6.0& 5.36& --     & -6.12   &  --6.14  &  --1.36  &  --1.38\\
AndXXIII    & 0.73& Sph-X& --9.5& 6.75& --     & -6.05   &     --    & --2.67  &    --\\
AndXXIV     & 0.60& Sph-X& --7.0& 5.77& --     &$<-$6.60   &    --    & $<-$2.23  &    --\\
AndXXV      & 0.81& Sph-X& --9.1& 6.58& --     &$<-$6.38   & $<-$6.28  & $<-$2.83  & $<-$2.73\\
AndXXVI     & 0.76& Sph-X& --6.5& 5.54 &--    & $<-$6.36   & $<-$6.30   &$<-$1.77  & $<-$1.71\\
AndXXVII    & 0.83& Sph-X& --7.3& 5.88& --    & $<-$6.42   & $<-$6.32  & $<-$2.18   & $<-$2.08\\
AndXXVIII   & 0.65& Tr-L & --7.7& 6.04& --    & $<-$6.60   &    -     &$<-$2.51    &  --\\
AndXXIX     & 0.73& Sph-X& --7.5& 5.96& --    &  -6.54   & $<-$6.54   & --2.36   &$<-$2.36\\
AndXXX      & 0.68&  --   &   -- &   - & --    &    -      &   --     &   --     &  --\\
Bol520      & 0.63& dE-H & --8.1& 5.60& --    & $<-$6.77    &--6.09    &$<-$2.24  &  --1.56\\
\hline
Median      & 0.76& --    & --7.5& 5.96& <5.54& $<-$6.38   &$<-$6.24   & $<-$2.36  & $<-$1.87\\ \hline
\end{tabular}
\end{table}

References to sources of the data are contained in the Local volume galaxies database (Kaisina et al. 2012), which is available
on the internet at http://www.sao.ru/lv/lvgdb.  It can be seen that the absolute magnitudes of all the satellites of
Andromeda and almost all the satellites of M81 lie within the interval [$-10^m > M_B > -5^m$].  As a comparison, we have
collected published data on satellites of the Milky Way (MW) with luminosities in the same range. The parameters
of 11 such satellites of our Galaxy are summarized in Table 3, which has the same structure as the first two tables.
Unfortunately, a measurement of the $H\alpha$  flux is given there for only one satellite, Leo T. No attempts were made to
determine $F(H\alpha)$   for the other MW satellites because of the large angular sizes of these objects.

It should be noted that near the minimum values of the $FUV$ fluxes there is a substantial uncertainty in the
value because faint blue stars of our Galaxy are projected onto the profile of a diffuse galaxy.  In several cases, these
values of the $FUV$ fluxes can  only serve as  upper limits.

\begin{table}
\caption{Integral Parameters of the Observed Satellites of M81}
\begin{tabular}{|c|c|l|r|c|c|r|r|r|r|}  \hline

Satellite&$D$, Mpc&  Type& $M_B$ &$\log M_*$& $\log M_{HI}$&  $\log SFR_{H\alpha}$& $\log SFR_{FUV}$ & $P(H\alpha)$ &$P(FUV)$\\
\hline
d0926+70 &3.93 &Tr-L & --10.0& 6.24  & $<$5.64 &$<-$5.12  & --4.18& $<-$1.23  & --0.29\\
d0934+70& 3.66 &Sph-X &--9.6& 6.80    &   --  &$<-$4.91  & --4.16 &$<-$1.59  & --0.84\\
d0939+71 &3.63& Tr-L & --8.4 &5.60    &   --  &$<-$5.48  & --5.79 &$<-$0.95  & --1.26\\
d0944+69& 3.98 &Sph-X &--7.4& 5.92    &    --& $<-$5.15  &$<-$4.90 &$<-$0.84 & $<-$0.59\\
d0944+71& 3.39 &Tr-L & --11.1& 7.39   &$<$5.49& $<-$5.15  & --5.74 &$<-$2.41  & --3.00\\
d0955+70& 3.93 &Sph-X& --9.1 &6.60   &     --& $<-$5.12  &$<-$4.74 &$<-$1.59  & $<-$1.21\\
d0959+68& 3.60& Ir-L & --10.1& 6.29  &    --  & -3.83  & --3.80 &   0.01 &   0.04\\
d1006+67& 3.87 &Sph-X&  --8.6& 6.37  &     --& $<-$5.27  &$<-$5.01 &$<-$1.51 & $<-$1.25\\
d1009+70 &9.00 &Sph-L& --12.4 & 7.90 &      -- &$<-$4.36 &  --3.29& $<-$2.12 &  --1.05\\
d1014+68& 4.57& Tr-X & --8.3& 6.27  & $<$5.84& $<-$4.93   & --4.58 &$<-$1.07  & --0.72\\
d1015+69& 4.07& Tr-X & --7.8& 6.06   &      -- &$<-$5.19 &  --5.39& $<-$1.11 &  --1.31\\
d1019+69& 9.60& Ir-L & --12.1 &7.07  &     -- & -3.51  & --3.31 & -0.44  &  --0.24\\
d1041+70& 3.91& Ir-X & --8.6 &5.71   & $<$5.28& $<-$5.03 & $<-$5.01& $<-$0.61 & $<-$0.59\\
\hline
Median& 3.93 &  --   &  --8.6& 6.27&   $<$5.56& $<-$5.12&  $<-$4.90& $<-$1.11&  $<-$0.84\\
\hline
\end{tabular}
\end{table}

\begin{table}
\caption{Integral Parameters of the Satellites of MW}
\begin{tabular}{|l|c|l|c|c|r|c|r|r|c|r|}  \hline
Satellite&$D$, Mpc&  Type& $M_B$ &$\log M_*$& $\log M_{HI}$&  $\log SFR_{H\alpha}$& $\log SFR_{FUV}$ & $P(H\alpha)$ &$P(FUV)$\\
\hline
Phoenix & 0.44 &Tr-L & --9.6& 6.08 &$<$5.22 &--        & --4.43& --      & --0.37\\
Carina  & 0.10 &Sph-X& --9.0 &6.54 &$<$2.32 &--         &--7.58& --      & --3.98\\
Leo T   & 0.42 &Ir-L & --6.7 &4.94 &  5.63& $<-$5.92  & --5.69& $<-$0.73 &  --0.50\\
Sex Sph & 0.09 &Sph-X& --8.7 &6.42 &$<$2.30 &--        &  --7.77& --    &  --4.06\\
CVn I   & 0.22 &Sph-X& --7.9 &6.10 &$<$4.50& --        &  $<-$7.67& --    &$<-$3.63\\
Boo III & 0.05 &Sph-X& -5.8 &5.29 &  --  & --        &   $<-$9.00& --    &  $<-$4.16\\
Boo I   & 0.07& Sph-X& --5.5 &5.15 &$<$2.01& --        &   $<-$8.73& --    & $<-$3.74\\
UMin    & 0.06 &Sph-X& --7.1 &5.80& $<$4.52& --        & $<-$8.76& --      &$<-$4.43\\
Herc    & 0.15& Sph-X& --6.1& 5.39& $<$2.72& --        &  $<-$7.88&--     &$<-$3.14\\
Draco   & 0.08 &Sph-X& --8.7& 6.45&$ <$2.19& --        &  $<-$8.52 &--    &  $<-$4.83\\
Tucana &  0.88 &Tr-L & --9.2& 6.62& $<$4.18& --        &     --6.0& --   & --2.52\\
\hline
Median &   0.10&   -- &--7.9& 6.08&  $<$2.72& --        &    $<-$7.77& --&$<-$3.74\\ \hline
\end{tabular}
\end{table}

\section{Discussion and conclusions}

The last row in Tables 1--3 characterizes the medians of the various parameters for satellites of the three nearest
massive spiral galaxies.  For samples of objects with highly different morphologies and fluxes, the medians are more
appropriate comparative characteristics than the means.  Contrasting these data, we can reach the following conclusions:

A.  The three samples of faint satellites around M81  n = 11), M31  n = 21), and MW  (n = 11) have similar
median absolute magnitudes, --8$\fm$6, --7$\fm$5, --7$\fm$9, and logarithms of the stellar masses,  6.27, 5.96, 6.08,  despite the
inevitable selection effects owing to the difference of a factor of 40 in the mean distances.

B. The medians of the upper limit on the logarithm of the hydrogen mass, $<$5.56 (M 81), $<$5.54 (M 31), and
$<$2.72 (MW), show that the low neutral hydrogen content in the spheroidal satellites of M81 and M31 may actually
be 2--3 orders of magnitude lower than the observed limit, by analogy with the satellites of our Galaxy.  The proportion
of hydrogen and stellar mass, $\log(M_{HI}/M_*$), for the satellites of M81 ($<-$0.71) and of M31 ($<-$0.42) should obviously
be closer to that characterizing the more studied satellites of the Milky Way ($<-$3.36).  As they pass through the dense
regions of neighboring giant spirals, the spheroidal dwarfs lose  almost all their reserves of gas needed for star
formation.

C.  On comparing the upper limits on the $SFR$ derived from the $H\alpha$  and $FUV$ fluxes, we can see that these
$H\alpha$  observations place a limit on the magnitude of the $SFR$ that is 2--3 times stricter than that for the $UV$ fluxes from
the GALEX satellite.  Here the observations in the $H\alpha$  emission line have 3--5 times higher angular resolution.

D.  Because of the evident selection effect, the median for the upper limit on the specific star formation rate,
$P(FUV$), decreases from distant to near subsystems: $<  -$0.84 (M 81), $<  -$1.87 (M 31), and $<  -3.74$ (MW).  However, the
three nonspheroidal satellites of the Milky Way have fairly high observed specific star formation rates: $-0.37$ (Phoenix),
$-0.50$ (Leo T), and $-2.52$ (Tucana).  These dwarfs, which are the most distant from the MW, have not yet lost their
gas components.

E.  Given the above mentioned trend with distance owing to the limits on the $H\alpha$  and $FUV$ flux measurements,
the actual specific star formation rate for spheroidal satellites of massive spiral galaxies are typically  $P <$ [ --3,--4].  In
other words, in order to reproduce the observed mass of  the spheroidal dwarfs over the cosmological time, their rates
of star formation in the past must have been $10^{(3-4)}$ times greater than at present.  Since the bulk of the stars in them
have ages in excess of 10 billion years, the star formation phase in the spheroidal dwarfs was extremely active in the
early era ( $z  > 10$).

F.  For most of the low-mass satellites of M31 and M81, only upper estimates are available for the $H\alpha$  and
$FUV$ fluxes.  Only four of the dwarf galaxies (d0959+68, And XI, And XVII, and And XXII) have $SFR$ values at a
level of $10^4-10^6 [M_{\odot}$/year]  obtained by both methods.  The agreement between these independent estimates (with
a spread of only 25\%) is quite unexpected and requires further analysis.  According to the data of Pflamm-Altenburg et al.  (2007, 2009),
estimates of the $SFR$ based on $H\alpha$  and $FUV$ fluxes can differ systematically by tens or hundreds of times when the
dwarf luminosities are extremely low.

In conclusion, we note the curious fact that the distant spherical cluster Bol520 of Andromeda has a low, but
fully determined, $FUV$ flux corresponding to log[$SFR$] = - 6.09. It is possible that, in accordance with current ideas,
the slow star formation occurring in these objects is caused by slow inleakage of the intergalactic medium that contains
the bulk of the baryonic component of the universe.

This work was supported by the Russian Foundation for Basic Research (RFFI), grant Nos. 13-02-90407, 12-
02-91338,  and  13-02-00780.

{\bf REFERENCES}

1. I. D. Karachentsev, S. S. Kaisin, Z. Tsvetanov, and H. Ford, Astron. Astrophys.  434, 935 (2005).

2. S. S. Kaisin and I. D. Karachentsev, Astrofizika  49, 337 (2006).

3. I. D. Karachentsev and S. S. Kaisin, Astron. J.  133, 1883 (2007).

4. S. S. Kaisin and I. D. Karachentsev, Astron. Astrophys.  479, 603 (2008).

5. I. D. Karachentsev and S. S. Kaisin, Astron. J.  140, 1241 (2010).

6. S. S. Kaisin, I. D. Karachentsev, and I. D. Kaisina, Astrofizika  54, 353 (2011).

7. E. I. Kaisina, D. I. Makarov, I. D. Karachentsev, and S. S. Kaisin, AstBul  67, 115 (2012).

8. I. D. Karachentsev, D. I. Makarov, and E. I. Kaisina, Astron. J.  145, 101 (2013).

9. A. W. McConnachie, Astron. J.  144, 4 (2012).

10. A. W. McConnachie and M. Irwin, Mon. Notic. Roy. Astron. Soc.  365, 1263 (2006).

11. K. Chiboucas, I. D. Karachentsev, and R. B. Tully, Astron. J.  137, 3009 (2009).

12. K. Chiboucas, B. A. Jacobs, R. B. Tully, and I. D. Karachentsev, Astron. J., accepted (2013).

13. I. D. Karachentsev, E. I. Kaisina, S. S. Kaisin, and L. N. Makarova, Mon. Notic. Roy. Astron. Soc.  415L, 31 (2011).

14. V. L. Afanasiev, E. B. Gazhur, S. R. Zhelenkov, and A. V. Moiseev, Bull. SAO  58, 90 (2005).

15. J. B. Oke, Astron. J.  99, 1621 (1990).

16. D. I. Makarov, L. Makarova, M. Sharina, et al., Mon. Notic. Roy. Astron. Soc.  425, 709 (2012).

17. S. Roychowdhury, J. N. Chengalur, K. Chiboucas, I. D. Karachentsev, et al., Mon). Notic. Roy. Astron. Soc.  426, 665 (2012).

18. D. F. de Mello, et al., Astron. J.  135, 548 (2008).

19. D. J. Schlegel, D. P. Finkbeiner, and M. Davis, Astrophys. J.  500, 525 (1998).

20. R. C. Kennicutt, ARA\&A  36, 189 (1998).

21. A. Gil de Paz, B. F. Madore, and O. Pevunova, Astrophys. J. Suppl. Ser.  147, 29 (2003).

22. A. Gil de Paz, S. Boissier, B. F. Madore, et al., Astrophys. J. Suppl. Ser.  173, 185 (2007).

24. J. Pflamm-Altenburg, C. Weidner, and P. Kroupa, Mon. Notic. Roy. Astron. Soc.  395, 394 (2009).

25. J. Pflamm-Altenburg, C. Weidner, and P. Kroupa, Astrophys. J.  671, 1550 (2007).

\clearpage

\begin{figure}
\includegraphics[scale=0.55]{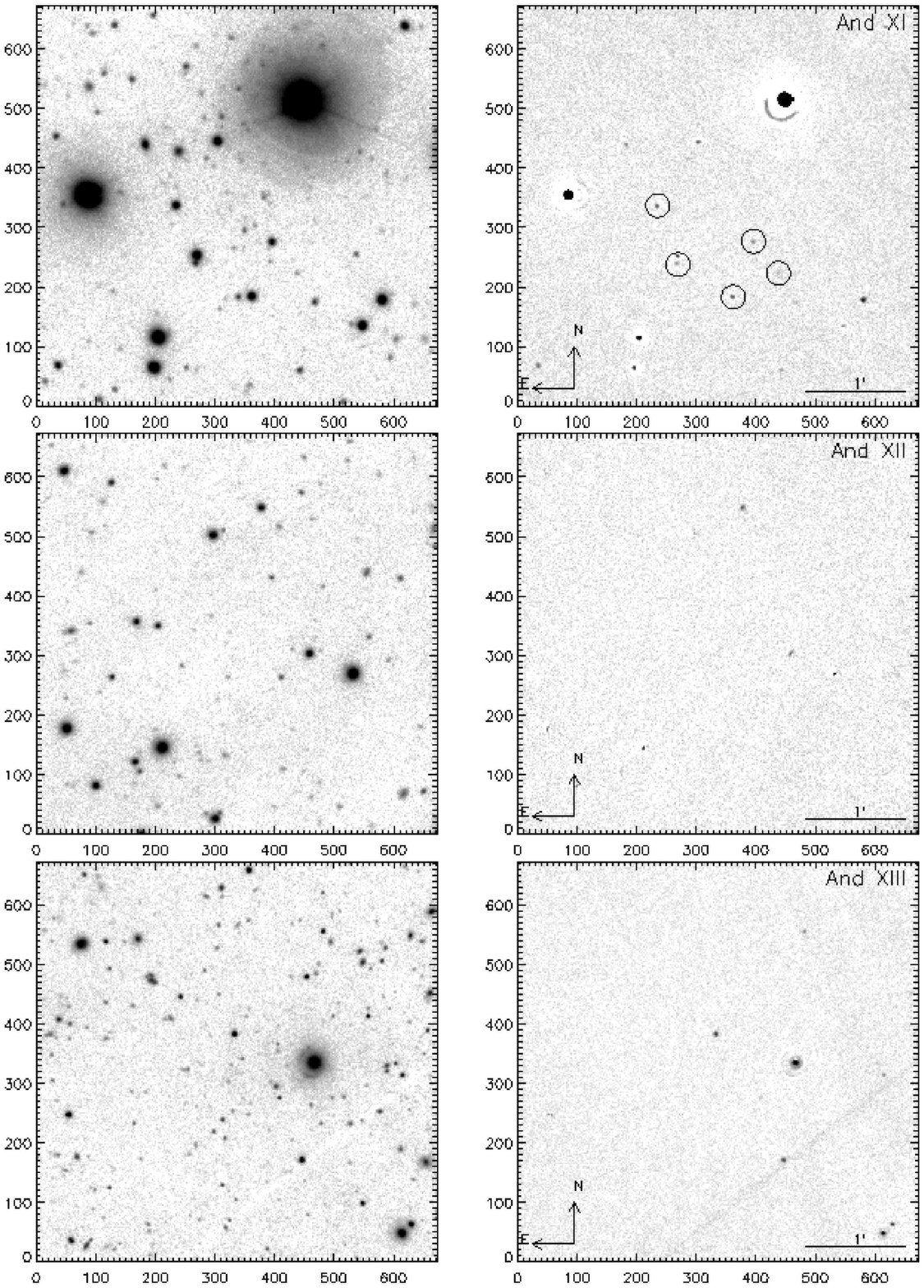}
\caption{Atlas of images of the satellites of M31}
\end{figure}

\setcounter{figure}{0}
\begin{figure}
\includegraphics[scale=0.55]{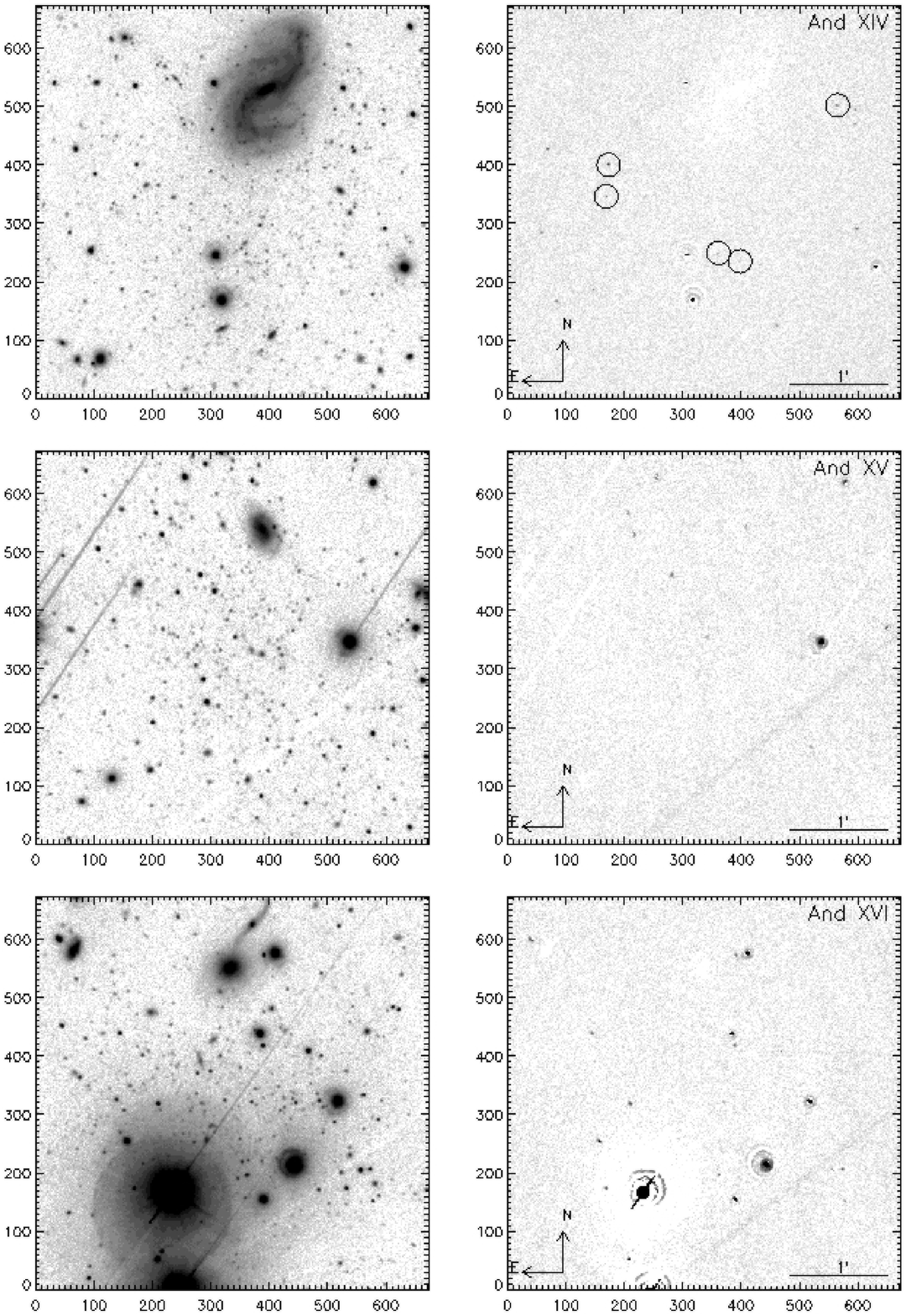}
\caption{Continued}
\end{figure}

\setcounter{figure}{0}
\begin{figure}
\includegraphics[scale=0.55]{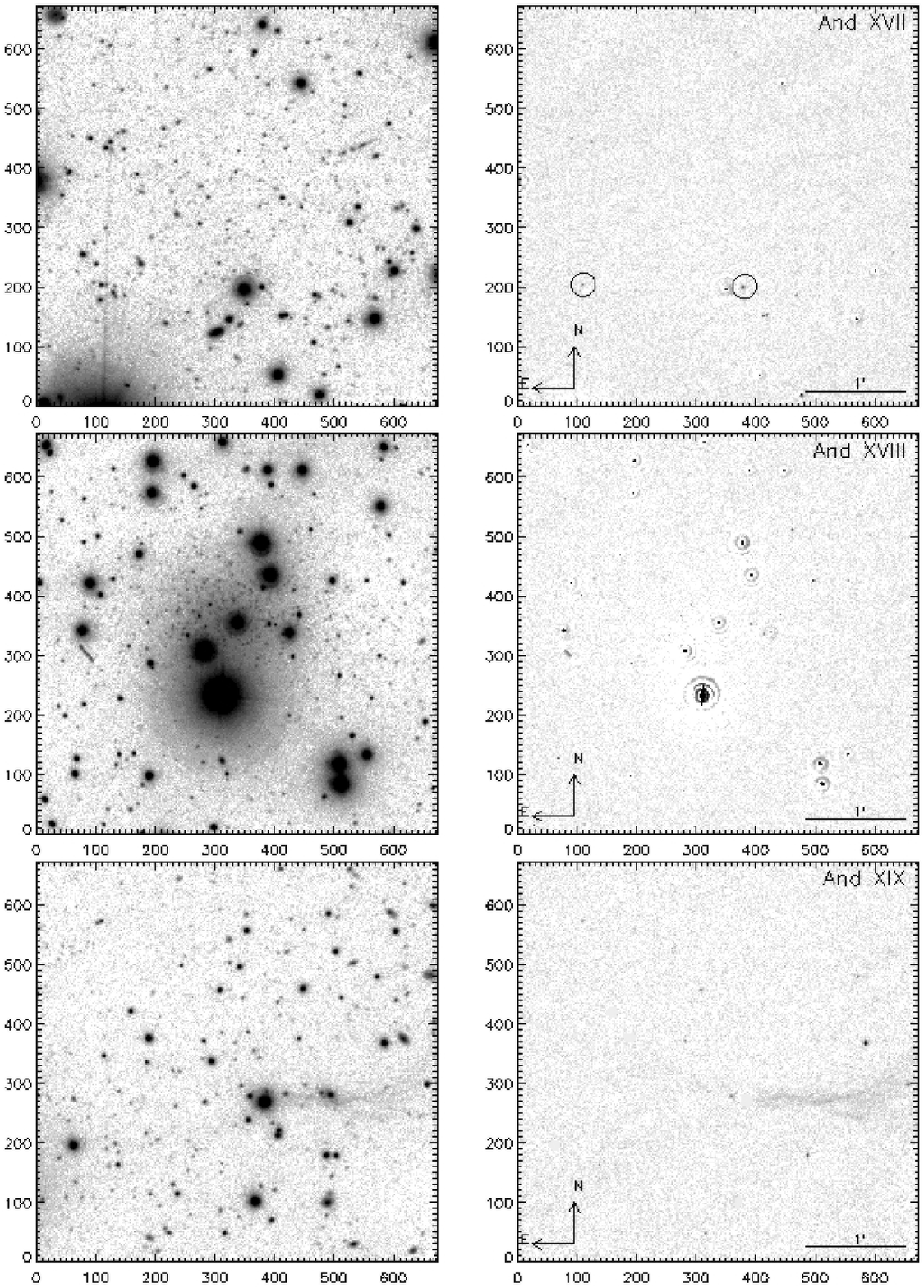}
\caption{Continued}
\end{figure}

\setcounter{figure}{0}
\begin{figure}
\includegraphics[scale=0.55]{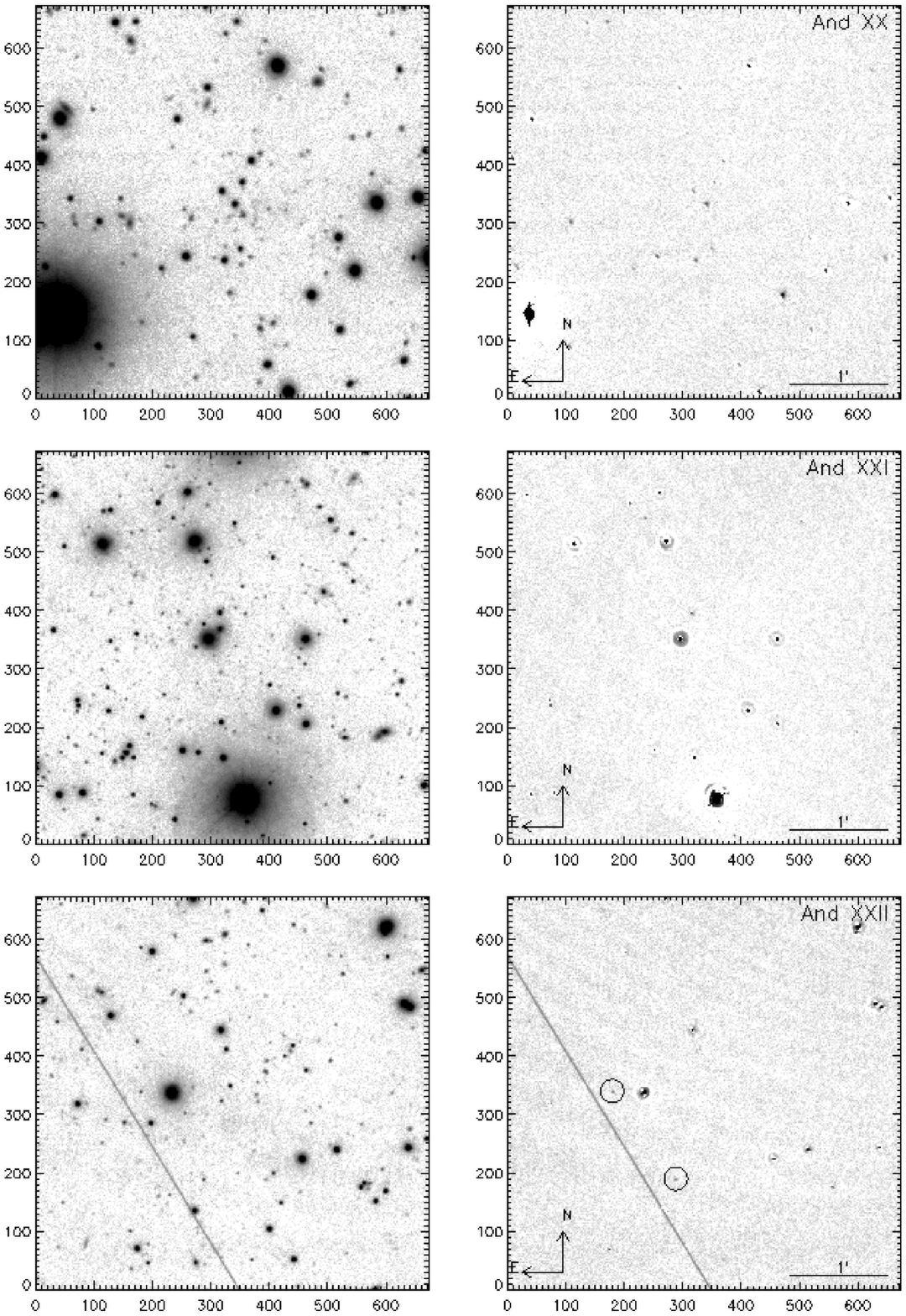}
\caption{Continued}
\end{figure}

\setcounter{figure}{0}
\begin{figure}
\includegraphics[scale=0.55]{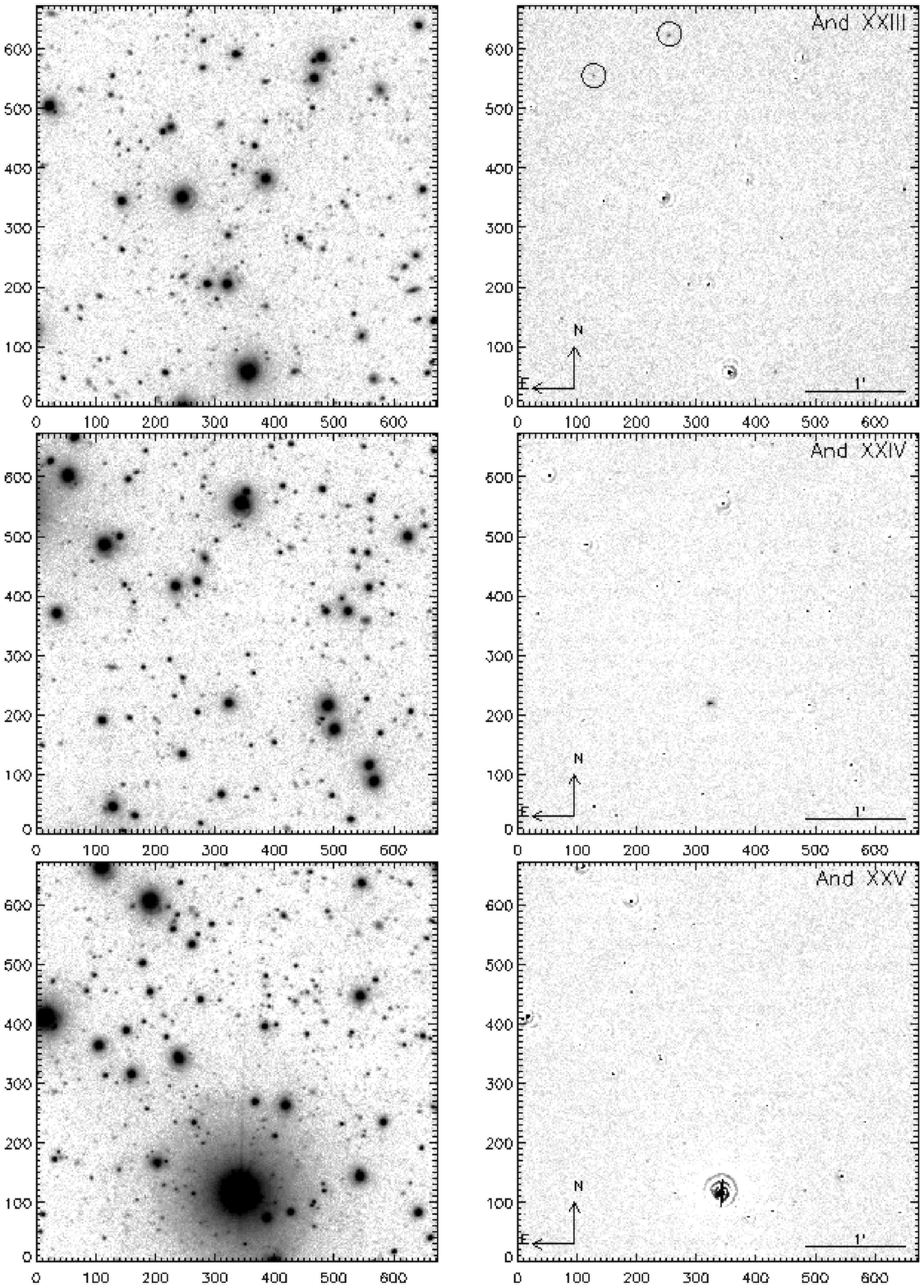}
\caption{Continued}
\end{figure}

\setcounter{figure}{0}
\begin{figure}
\includegraphics[scale=0.55]{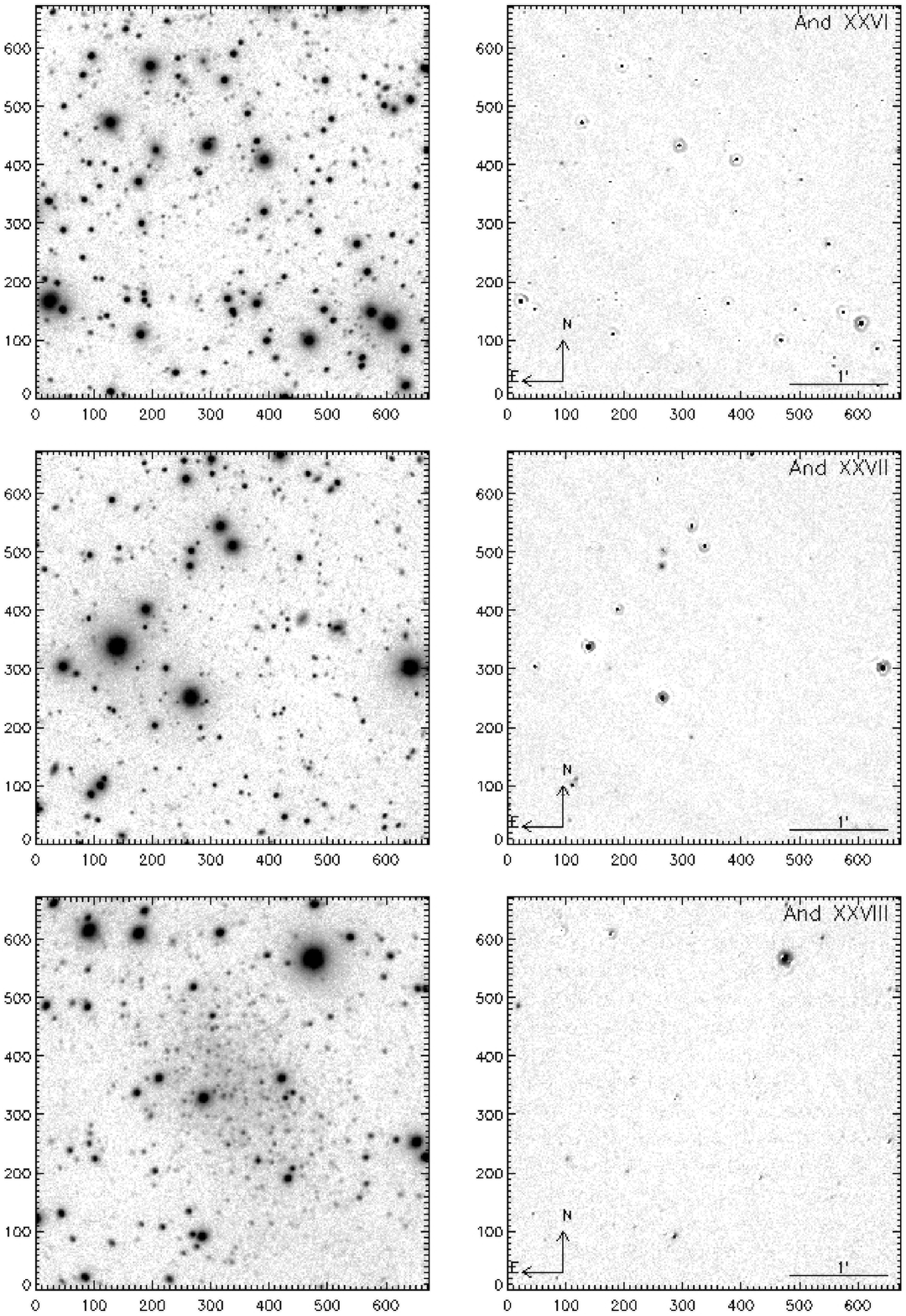}
\caption{Continued}
\end{figure}

\setcounter{figure}{0}
\begin{figure}
\includegraphics[scale=0.55]{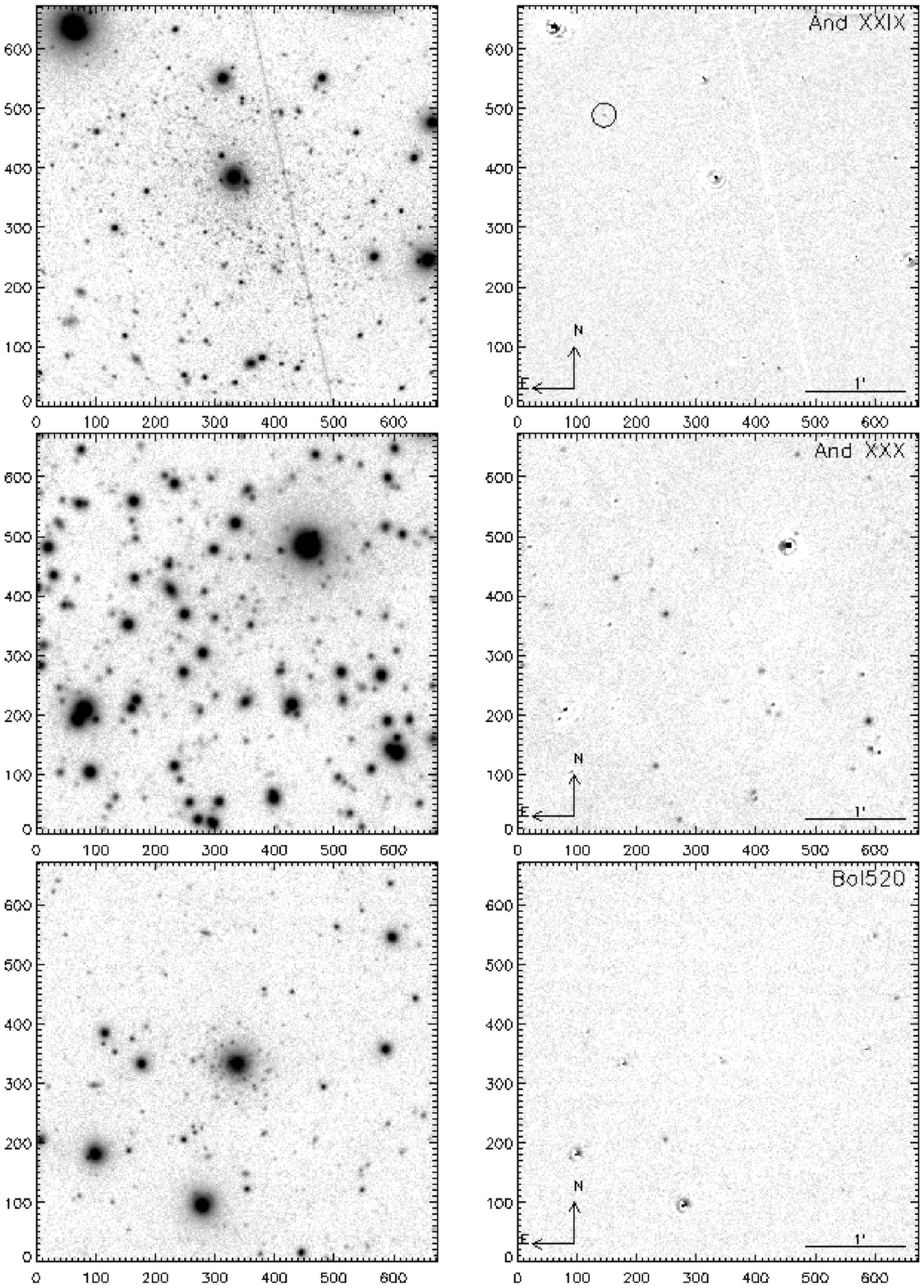}
\caption{Conclusion}
\end{figure}

\begin{figure}
\includegraphics[scale=0.55]{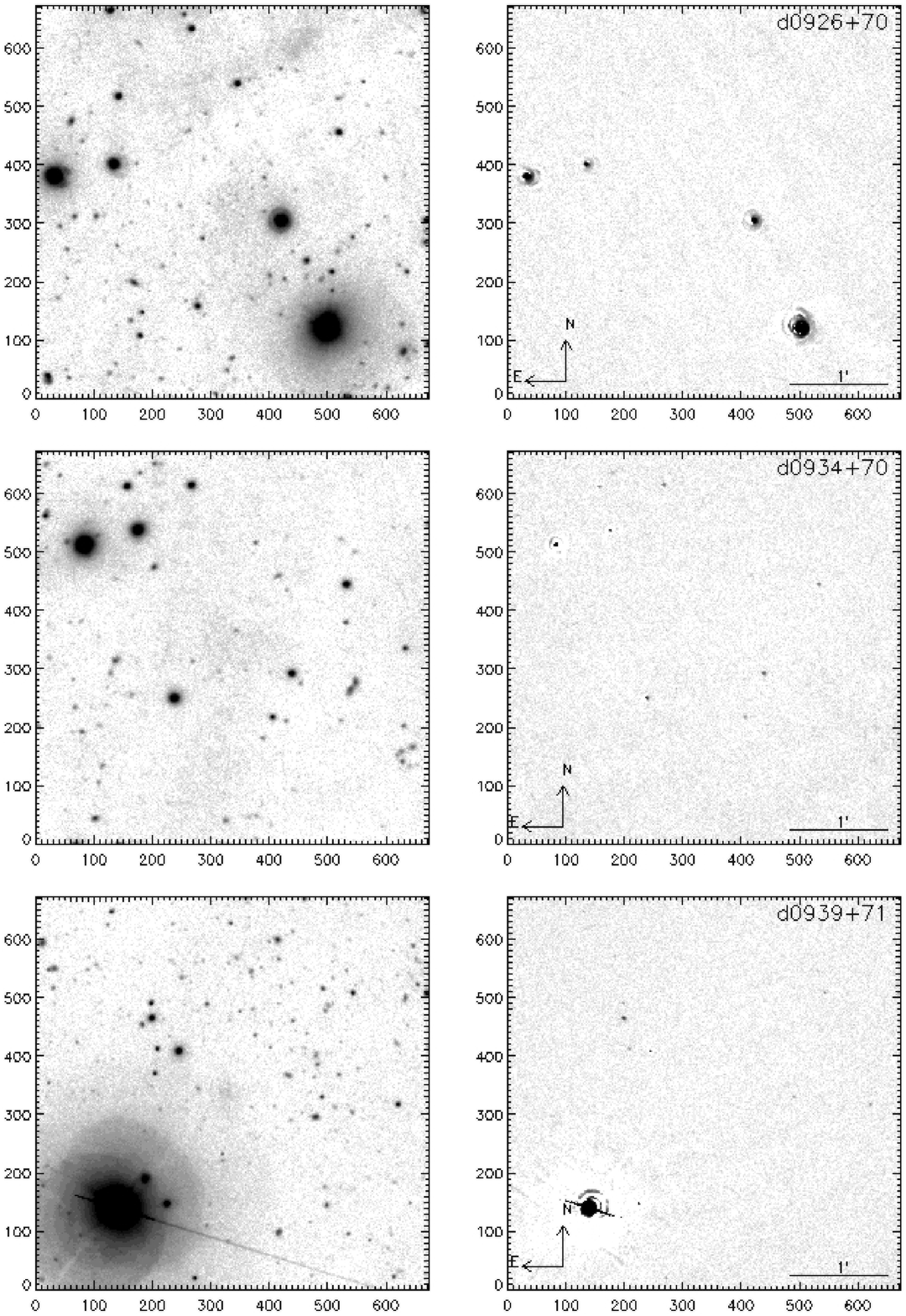}
\caption{Atlas of images of the satellites of M81}
\end{figure}

\setcounter{figure}{1}
\begin{figure}
\includegraphics[scale=0.55]{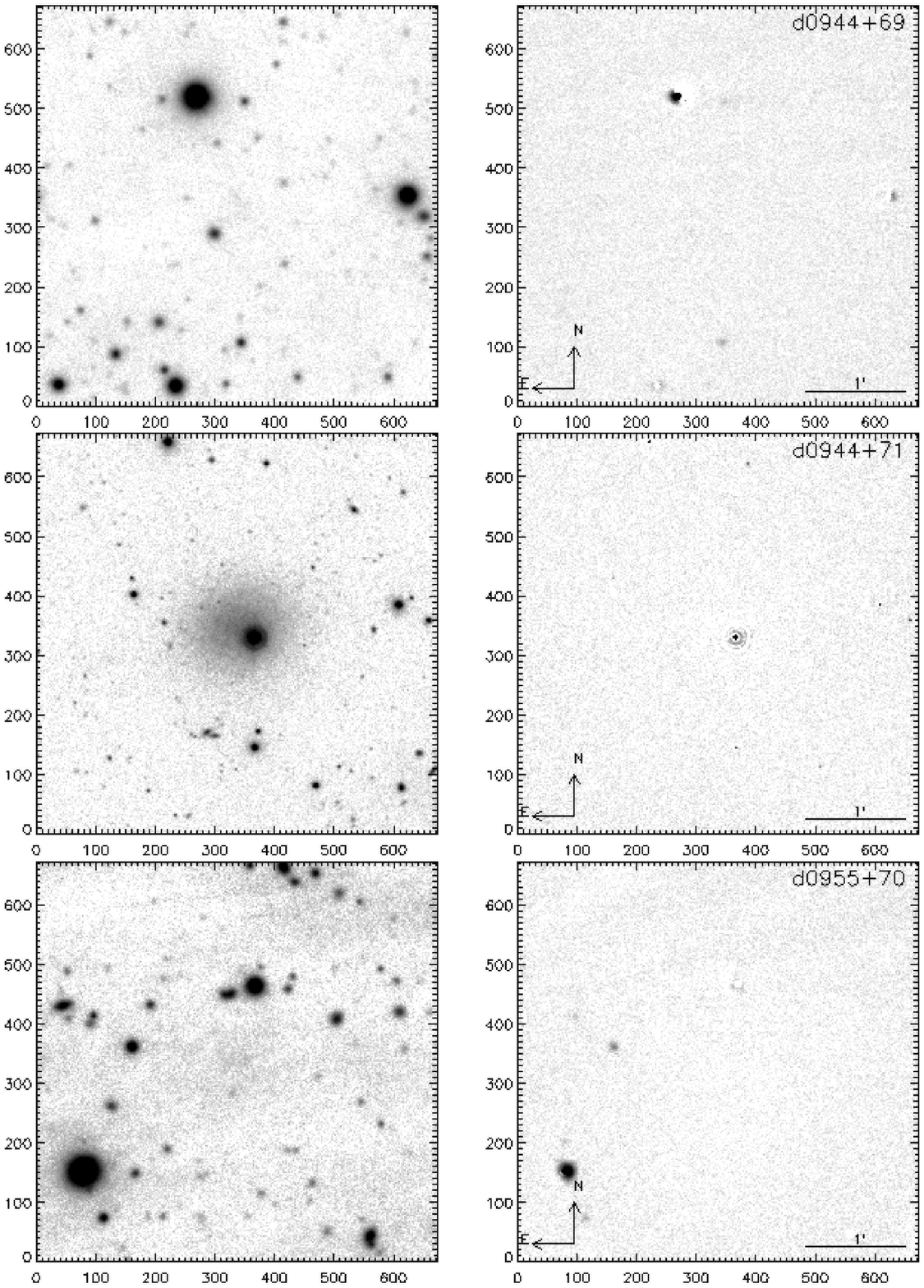}
\caption{Continued}
\end{figure}

\setcounter{figure}{1}
\begin{figure}
\includegraphics[scale=0.55]{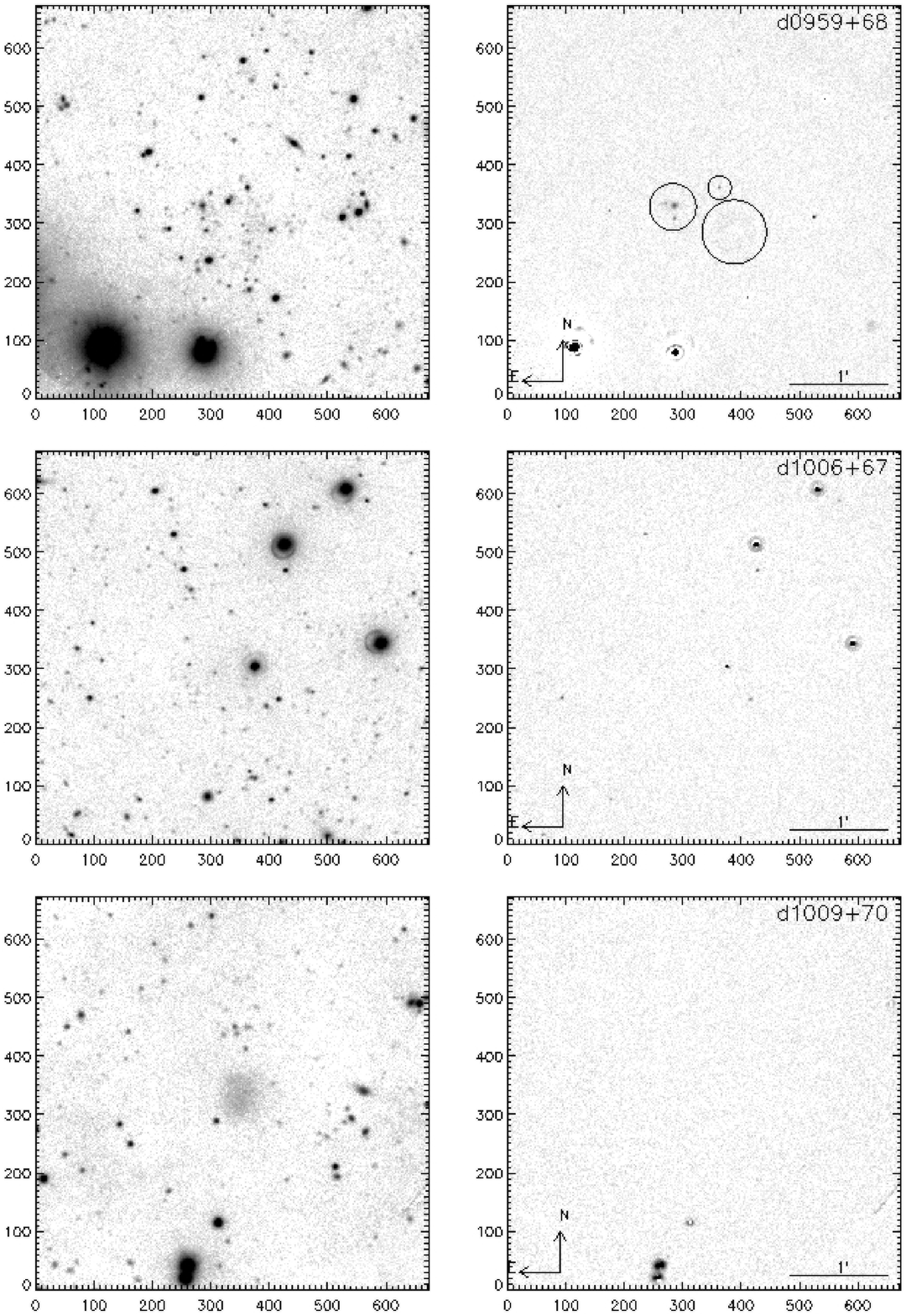}
\caption{Continued}
\end{figure}

\setcounter{figure}{1}
\begin{figure}
\includegraphics[scale=0.55]{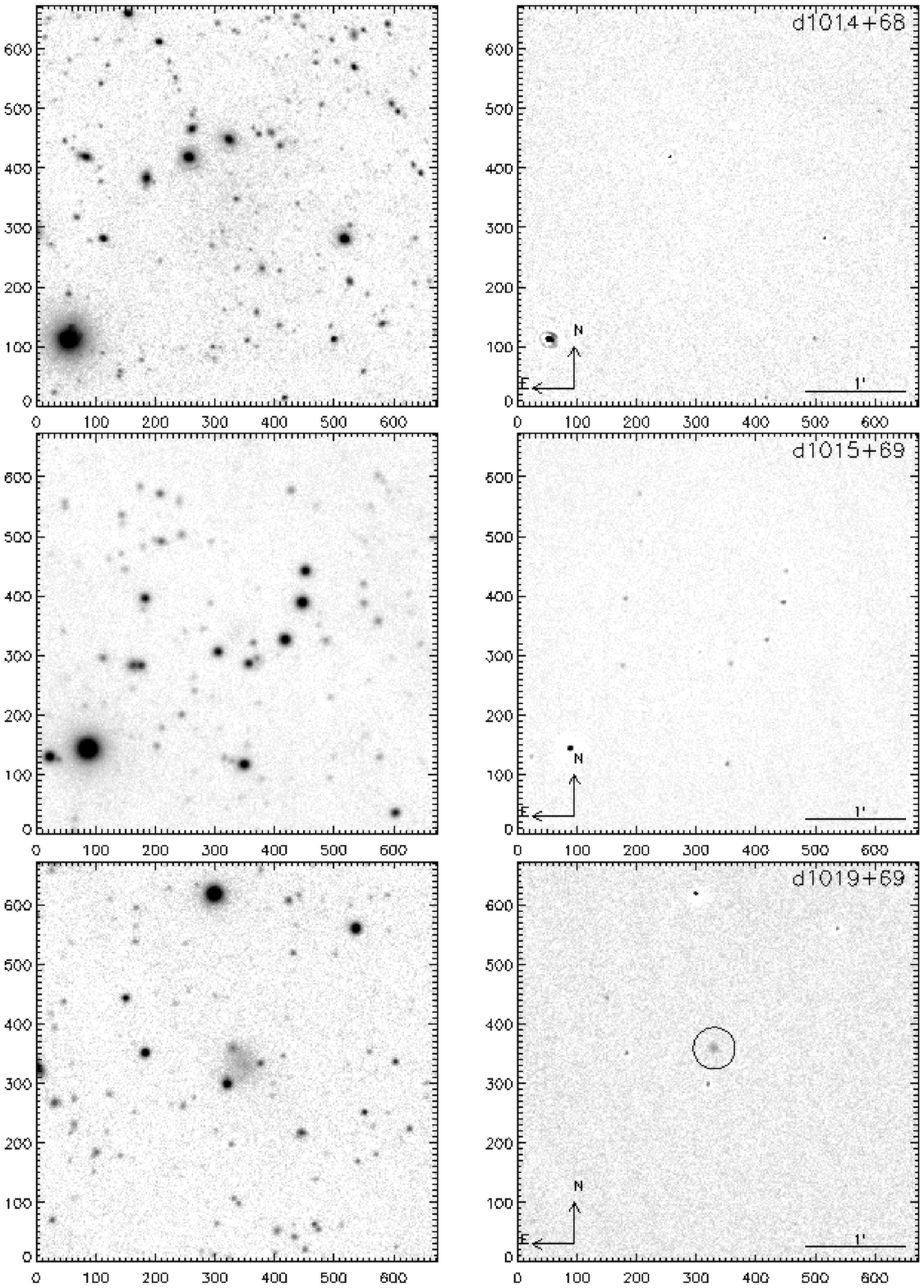}
\caption{Continued}
\end{figure}

\setcounter{figure}{1}
\begin{figure}
\includegraphics[scale=0.55]{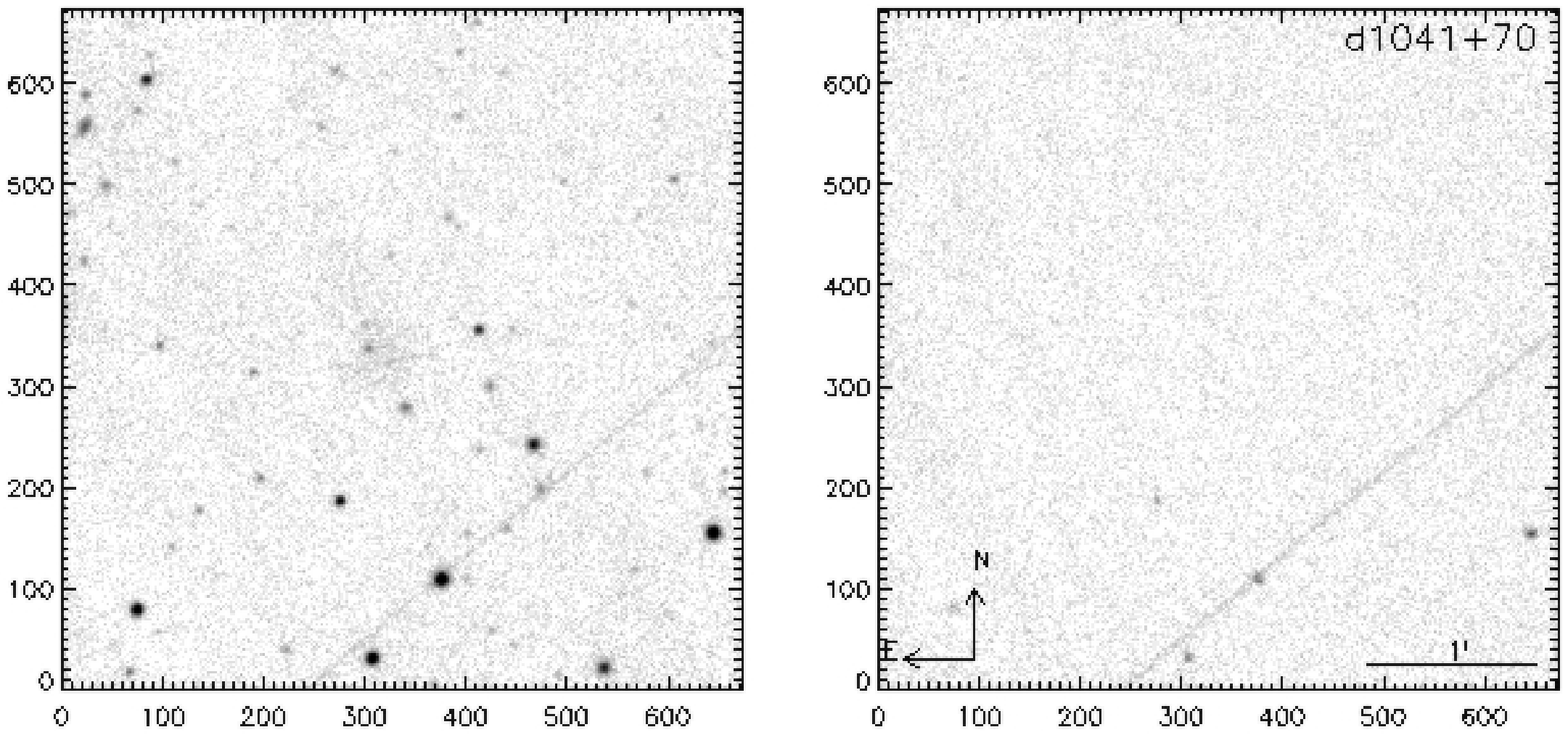}
\caption{Conclusion}
\end{figure}

\end{document}